%% file: paper.tex
\documentclass{article}
\PassOptionsToPackage{numbers,sort&compress}{natbib}
\usepackage[preprint]{tmlr}
\setcitestyle{numbers,square,citesep={,}}
\usepackage{amsmath,amssymb}
\usepackage{booktabs}
\usepackage{graphicx}
\usepackage{microtype}
\usepackage{url}
\DeclareUrlCommand\doi{\urlstyle{rm}}
\providecommand{\href}[2]{#2}
\newcommand{\artifacturl}{https://github.com/rodriguescarson/cad-silent-failure-bench}
\newcommand{\artifactdoi}{10.5281/zenodo.22312792}

\title{\raggedright Done Is Not Correct: Measuring Silent Failures and Self-Verification
Calibration When LLM Agents Take CAD Actions}

\author{\name Carson Rodrigues \email carson@celabe.com \\
        \addr Celabe
        \AND
        \name Clive Rodrigues \email rodriguesclive8@gmail.com \\
        \addr Hochschule Coburg, University of Applied Sciences
        \AND
        \name Aravind Reddy G \email aravind99.g@gmail.com \\
        \addr Independent}

\def\month{07}\def\year{2026}\def\openreview{\url{https://openreview.net/forum?id=XXXXXXXX}}

\begin{document}
\maketitle

\begin{abstract}
When a large language model (LLM) agent \emph{builds} a mechanical part by writing parametric CAD
code from a natural-language specification, syntactic success (the code runs, the geometry
renders) decouples from semantic correctness (right dimensions, valid constraints, manufacturable
features). We release a benchmark and harness that measure this gap directly: 13 specification
tasks in three tiers whose every requirement maps to machine-checkable properties of the produced
B-rep (dimensions within published tolerance bands, feature counts with axial extents, per-hole
bolt-circle positions, mass-centre offsets, volume), measured exactly by the kernel; reference
solutions and a calibration gate; and the full attempt corpus, with final code, measured
properties and verdicts for every graded attempt, full transcripts for every fixed-protocol
tool-use and ablation attempt, and a 3D viewer. The exact, self-audited oracle isolates two quantities no existing CAD
benchmark reports: the \textbf{silent-failure rate} (a valid, renderable solid, an explicit
completion claim, and at least one failed semantic check) and \textbf{completion-claim
calibration}. Across four models from four vendors ($n{=}156$ attempts per architecture over 13
tasks, cluster-robust intervals, paired tests), single-shot agents claimed completion on 98\% of
attempts and were silently wrong on 14.1\% (task-cluster CI [5.8, 24.4]); a tool-use loop that
enforces \emph{measure-then-claim} cut silent failures to 3.2\% (McNemar $p{<}10^{-4}$) and raised
claim precision from 64\% to 96\%. A single-variable ablation on all four models attributes the change to the claim contract rather
than to the feedback: with the legacy contract and identical feedback, silent failures were 14 of
156 against 5, and claim precision 68\% against 96\%. A planner--coder--checker pipeline matched that precision at
lower coverage with no success gain. Verbalized confidence was poorly calibrated single-shot (ECE
0.32) and became informative under verification for three of four models. Context divergence
between roles predicted episode breakdown (AUROC 0.86) no better than the number of rounds did
(0.87), and did not flag valid-but-wrong parts. The benchmark audits itself: hardening our own
oracle flipped 18 of the 519 verdicts it regraded from pass to fail and none in reverse, exposing
17 of that corpus's 48 silent failures, and a blind expert audit overturned 3 of the 18 primary
silent failures outside the one task whose specification admits two readings, so the rates are
bounded on both sides. Scaling every tolerance band from a quarter to four times its published width
changes no verdict in the verified conditions above half-width and leaves the single-shot to
tool-use contrast intact (McNemar $p < 0.01$ at every scaling), so the conclusion does not rest
on where the bands were drawn.
\end{abstract}

\section{Introduction}
LLM agents increasingly act rather than advise: they write and execute code, call tools, and report
back \citep{yao2023react, schick2023toolformer}. Software engineering already has
execution-grounded benchmarks for this regime \citep{jimenez2024swebench}, and a growing body of
evidence that agent evaluation systematically overstates reliability: graders admit false positives
\citep{yu2025utboost, zhu2025benchpractices}, headline accuracy hides inconsistency
\citep{yao2024taubench, kapoor2024aiagents}, web-agent progress shrinks under rigorous re-evaluation
\citep{xue2025illusion}, and agents asked to estimate their own success are grossly overconfident
\citep{kaddour2026overconfidence}.

Mechanical computer-aided design (CAD) is the next domain in this trajectory; it is also, we
argue, the best instrumented one for studying agent reliability. A wave of recent work generates parametric
CAD from text \citep{khan2024text2cad, xie2025texttocadquery, wang2026text2cadbench}, increasingly
with agentic feedback loops over a CAD kernel \citep{mallis2024cadassistant, yuan2026procad,
barkley2026cadsmith, ocker2025ideatocad}, and surveys now map the area \citep{zhang2025llmcadsurvey}.
Yet evaluation in this literature measures \emph{generation quality}: Chamfer distance or IoU to a
reference solid, execution success, or judge scores \citep{zhou2025cadjudge,
alrashedy2024cadcodeverify}. None of it asks the reliability question that determines whether an
engineer can delegate: \emph{when the agent says the part is done, is it, and does the agent know?}

Crashes are cheap to catch: a spec that fails to execute announces itself. The expensive
failure is a part that builds and renders plausibly while its bolt circle sits on the wrong pitch
diameter, delivered with ``done, confidence 0.97.'' Companion work by the first author \citep{rodrigues2026contextdrift}
defines context drift in multi-agent LLM pipelines, a divergence between agents' working
beliefs that we also measure here (Section~\ref{sec:method}); people relying on AI advice often
fail to rely on it appropriately \citep{schemmer2023appropriate}; and the self-correction literature explains why the agent's own check is not to be trusted
without external feedback \citep{huang2024selfcorrect, kamoi2024selfcorrection, gou2024critic}. CAD
supplies what those settings lacked: exact measurement. The produced B-rep can be measured
(dimensions, feature counts and extents, hole positions, mass properties, volume), so agent-side
overclaiming can be separated from grader inadequacy to a degree execution-graded software
benchmarks cannot \citep{yu2025utboost}; we audit and quantify the residual grader risk rather than assume it to be zero.

\paragraph{Contributions.}
(1)~A reproducible benchmark and harness: 13 natural-language mechanical specification tasks in
three tiers, each requirement mapped to machine-checkable properties of the produced B-rep with
published tolerance bands, with reference solutions and a calibration gate, built on an open kernel
(build123d/OpenCascade). It is released with every graded attempt (final code, measured properties,
verdicts under both oracle versions, transcripts for the fixed-protocol runs), the grader, the
harness, and a 3D viewer (Section~\ref{sec:availability}).
(2)~A \textbf{claim-conditioned, self-auditing exact-oracle silent-failure rate coupled to
completion-claim calibration}. The silent-failure rate is the conjunction of (i)~valid renderable
geometry, (ii)~an explicit agent completion claim, and (iii)~failure of at least one semantic
check, graded by an exact kernel oracle that we audit against ourselves; it is coupled to
completion-claim calibration measured as ECE, reliability diagrams, AUROC,
and Brier score \citep{guo2017calibration, tian2023justask, kaddour2026overconfidence}. Conditioning
on the agent's own claim and on a self-audited exact oracle is what separates this from the
generation- and repair-quality scores of CADSmith \citep{barkley2026cadsmith}, the finite-element
requirement-pass rate of \citet{son2026feacad}, and the assembly-structure, interface and
functional scores of MUSE \citep{dong2026muse}; we do not claim to be first to measure geometric correctness.
(3)~A controlled comparison of agent architectures (single-shot, tool-use with kernel measurement
feedback, and a planner--coder--checker pipeline with context-drift instrumentation), with a
single-variable ablation of the claim contract, testing whether feedback closes the
syntactic--semantic gap or merely hides it \citep{cemri2025mast, xie2026cascades}.
(4)~Findings with cluster-robust statistics over four model vendors: unverified completion claims
carry a 14.1\% silent-failure rate; enforcing measure-then-claim cuts it to 3.2\%, and the ablation
attributes the cut to the claim contract; an independent checker matches that claim precision at
lower coverage; stated confidence becomes informative under verification for three of four models;
context divergence predicts breakdown no better than episode length and misses subtle wrongness;
hardening our own oracle flipped 18 verdicts and none in reverse, so unmodelled properties can only
raise the rates, while the expert audit overturns 3 of the 18 silent failures outside the one
ambiguously specified task; and the measurand is stable under scaling of every tolerance band (F7).
We also report which of our pre-specified predictions the data did not support
(Section~\ref{sec:discussion}).

\section{Related work}
\paragraph{Text-to-CAD generation and benchmarks.}
CAD datasets of construction sequences, constrained sketches, B-rep models and labelled mechanical
parts \citep{wu2021deepcad, willis2021fusion, koch2019abc, seff2020sketchgraphs, kim2020mcb} enabled one-shot text-to-CAD models \citep{khan2024text2cad,
xie2025texttocadquery, guan2025cadcoder, xu2024cadmllm} and benchmarks scoring generation quality
across complexity tiers \citep{wang2026text2cadbench}. Graders exist as training rewards
\citep{zhou2025cadjudge} and VLM-based self-checks \citep{alrashedy2024cadcodeverify,
badagabettu2024query2cad}. Agentic systems
add kernel or visual feedback: tool-augmented VLLMs over FreeCAD \citep{mallis2024cadassistant},
clarification agents that resolve spec ambiguity \citep{yuan2026procad}, multi-agent design teams
\citep{ocker2025ideatocad}, agentic data synthesis at scale \citep{ataei2026zerotocad}, and
workflow comparisons graded by human visual inspection \citep{schuepbach2025texttodesign}.
CADSmith \citep{barkley2026cadsmith} is closest to our infrastructure: it validates agent-generated
CadQuery against exact kernel measurements (alongside a VLM judge) and built its own test prompts
because reference geometry in earlier LLM-generated datasets had ``silently diverged'' from prompt
intent, but it treats such divergence as a defect to engineer away. We treat it as the \emph{measurand}: our benchmark quantifies how often agents ship it while
claiming success, per architecture and per model. Two concurrent 2026 systems push geometric
grading further along different axes. The FEA-feedback agent of \citet{son2026feacad} assembles
multi-part STEP files from engineering briefs and validates them against typed physical and
geometric requirements with a finite-element solver; this scores a requirement-pass rate rather
than a claim-conditioned silent-failure rate, depends on a meshing and solver stack rather than a
self-auditing exact-geometry oracle, and has no calibration axis. MUSE \citep{dong2026muse} grades
multi-component B-rep assemblies, combining direct B-rep analysis of assembly structure and
interfaces with a VLM judge for functional consistency (a leaky oracle by the same argument we apply
to visual self-checks); it scores generation without an agent loop and reports no silent-failure
rate or completion-claim calibration. 3DCodeBench \citep{gao2026threedcodebench} benchmarks agents
that write procedural 3D code, but in the art-and-object domain with image- and shape-similarity
metrics and pairwise human preference, not mechanical tolerances or a silent-failure rate. Against all of these, our contribution is a
\emph{claim-conditioned, self-auditing exact-oracle silent-failure rate coupled to completion-claim
calibration}: we condition the metric on the agent's own completion claim, grade with an exact
kernel oracle we audit against ourselves (Section~\ref{sec:results}, F6), and report the
calibration of that claim, none of which these systems provide. We do not claim to be first to
measure geometric correctness; we claim a distinct measurand, graded by an exact, self-audited
oracle with no judge model in the loop. Topological property checks
\citep{preintner2025evocad} complement our dimensional and positional ones.

\paragraph{Agent reliability and overclaiming.}
Execution-grounded agent benchmarks measure end-state success \citep{jimenez2024swebench,
yao2024taubench}, but their oracles leak: insufficient tests mislabel hundreds of SWE-bench patches
as resolved \citep{yu2025utboost}, and systematic audits find task- and outcome-validity flaws
across agentic benchmarks \citep{zhu2025benchpractices, kapoor2024aiagents, xue2025illusion}.
Orchestration-level work names silent failures (wrong-but-plausible outputs with no error
signal) and engineers verifier layers against them \citep{babu2026selfhealing}; stress-testing
frameworks measure consistency under perturbation \citep{gupta2026reliabilitybench}. Multi-agent
failure taxonomies identify verification failure and false consensus as recurring modes
\citep{cemri2025mast, xie2026cascades}. Closest in construct, \citet{advani2026silentfailure}
characterize the same false-success phenomenon directly against environment-state ground truth:
it accounts for 45--48\% of failures in single-control $\tau^2$-bench domains and 75.8\% of
AppWorld coding-agent trajectories with explicit status claims, and no LLM-judge configuration
detects it above AUROC~0.65 on $\tau^2$-bench. Their ground truth is the end state of a task; ours
is the produced part measured against every stated requirement by an exact kernel, which lets us
grade each requirement separately, audit the oracle itself, and couple the claim to stated
confidence. Our silent-failure rate is conditioned on the \emph{agent's own completion claim} under
that self-audited geometric oracle, so it isolates agent-side overclaiming rather than grader
weakness or injected faults.

\paragraph{Self-verification and calibration.}
LLMs can verbalize confidence \citep{lin2022verbalized}; larger models are well calibrated on
suitably formatted questions and can estimate whether their own answers are correct
\citep{kadavath2022know}, and verbalized confidence beats token probabilities for RLHF models
\citep{tian2023justask}, but verbalized confidence tends to be overconfident
\citep{xiong2024uncertainty}. Intrinsic self-correction does not
work \citep{huang2024selfcorrect}; external, tool-grounded feedback does \citep{gou2024critic,
kamoi2024selfcorrection}. For code, confidence is poorly calibrated against execution-based
correctness \citep{spiess2025calibration}. Closest to us, \citet{kaddour2026overconfidence} measure
agents' predicted-versus-actual success on software tasks and find up to 55-point gaps, with
post-execution review discriminating successes from failures no better than pre-execution
estimates made with less information. We instantiate this question in a domain with an exact
geometric oracle, add the architecture axis, and couple calibration to a silent-failure rate;
concurrent work finds that LLM judges detect false success no better than AUROC~0.65
\citep{advani2026silentfailure}, and here no judge stands between the claim and the verdict,
because the kernel oracle grades every claim directly.

\section{The benchmark and harness}
\label{sec:method}

\paragraph{Task schema.}
\sloppy A task is data, not code: \texttt{\{id, tier, nl\_spec, units, checks[]\}}. Each check names a
\emph{kind} the grader knows how to evaluate against a measured-properties dictionary: scalar
comparisons with tolerance (\texttt{approx}, \texttt{range}, \texttt{equal}) over a fixed property
vocabulary (bounding-box extents, volume, centre of mass, solid/face/edge counts), boolean validity
checks (\texttt{valid\_solid}, \texttt{watertight}), and specialized geometric checks
(\texttt{cylinder\_count}, \texttt{bolt\_circle} with pitch-circle diameter and positional
tolerance). Specs are written so that \emph{every} requirement in the natural language maps to at
least one computable check; this discipline is what makes the suite auto-gradable and auditable.

\paragraph{Tiers and variation.}
The v1 suite has 13 tasks in three tiers: \textbf{T1} single feature (flange with bolt circle,
washer, bushing, four-hole plate, shaft collar), \textbf{T2} multi-feature (stepped shaft, flanged
bushing, counterbored plate, three-step pin, pulley blank, L-bracket with non-coaxial holes), and
\textbf{T4} edit-under-constraint (modify provided code: re-pattern a bolt circle; enlarge a bore
under a material-volume cap). The suite has no multi-part assembly tier: that
requires multi-solid measurement and is not part of this release. Every requirement is graded
against a tolerance band; bands are engineering-informed, published per check with the suite, and
their effect on the measurand is tested directly (F7). To our knowledge no existing
CAD benchmark grades against tolerance bands rather than exact values or mesh similarity. The suite
is small by design: 13 tasks and 122 checks are enough to separate agent architectures with paired
tests across four vendors, and the released schema is what a larger suite extends
(Section~\ref{sec:discussion}, roadmap).

\paragraph{Execution and measurement.}
Agents emit Python for the open \texttt{build123d} kernel (OpenCascade). Code runs in an isolated
subprocess that builds the solid and measures kernel-truth properties; a crash or hang cannot take
down the runner, and the only OpenCascade dependency is quarantined in that one module. Cylindrical
features are detected by grouping cylindrical faces by their infinite supporting cylinder (radius
$+$ axis line) and merging groups by \emph{axial extent}: overlapping spans are one physical
feature (a hole whose face a boolean split at the seam), disjoint spans are separate features
(coaxial bearing seats). The extractor is lenient on wrapper conventions (builder objects, lists of
solids) and strict on geometry.

\paragraph{Agent conditions.}
\textbf{Single-shot}: one completion from the spec. \textbf{Tool-use}: a ReAct-style loop
\citep{yao2023react} in which the harness executes each code revision and returns the measured
geometry of the solid: validity, bounding-box extents, volume, centre of mass, solid count, and,
for every detected cylindrical feature, its radius, axis, centre, and axial extent. The feedback
never includes the specification's targets, tolerance bands, or any pass/fail verdict: the agent
sees what a kernel reports, not what the grader concludes, and must still map the specification
onto those numbers itself. A completion claim that accompanies
fresh, unverified code does not end the episode; the agent must see kernel measurements for its
final code before its \textsc{done} is accepted, ensuring the condition actually delivers the
feedback it is named for. \textbf{Multi-agent}: planner$\to$coder$\to$checker, all three roles played by the same model, where the checker judges from the
identical kernel-measured property set (plus a watertightness flag) and never from the checks
themselves, so the two verified conditions are information-matched and differ only in
architecture. Each role ends its turn with a flat JSON of the governing dimensions it
believes (\texttt{DIMS}); the per-round \textbf{Context Divergence Score (CDS)} is the mean
pairwise disagreement over the union of stated dimension keys (a key counts as agreeing when
present for both roles and within 2\% relative tolerance; the planner's key vocabulary is
propagated so naming variation is not scored as drift). CDS $=0$ means perfectly aligned beliefs,
$1$ means none shared; it operationalizes context divergence for this domain and tests whether
verification-stage failures and false consensus \citep{cemri2025mast, xie2026cascades} predict
silent failures. In every condition the agent is
instructed to report \textsc{done} plus a verbalized confidence in $[0,1]$ when it believes all
requirements are met; verbalized elicitation follows \citet{tian2023justask}.

\paragraph{Grader soundness.}
Every task ships a reference solution (never shown to agents). A calibration gate requires each
reference solution to pass every one of its task's checks before any agent run is scored; a
stratified sample of 81 valid-solid agent outputs from the primary population (all 31 of its
silent failures, all 25 of its withheld-but-wrong parts, and 25 randomly drawn grader-passes) was
packaged with rendered geometry and blind-coded by a practising mechanical (CAD) engineer, a
co-author who wrote neither the specifications nor the grader, against the released
specifications, scoring each part pass/fail without sight of the grader verdict. Expert-versus-grader
agreement is \emph{substantial}: Cohen's $\kappa = 0.64$ (raw agreement $82.7\%$, $n=81$). Ten of the
fourteen disagreements trace to a single task, \textsc{t2-lbracket}, whose natural-language
specification admits two readings of the base depth (the $8$~mm wall inside the stated $60\times40$
footprint, giving an overall depth of $40$~mm, versus a full $60\times40$ base with the wall added
outboard, giving $48$~mm). Standard angle-bracket practice dimensions legs heel-to-toe, i.e.\ as
overall lengths, and our own wording fixes the vertical leg as an ``overall height,'' so the $40$~mm
base depth is likewise overall; the grader adopted this reading and the expert the other. This is a
finding about natural-language underspecification rather than a grader defect: excluding that one
ambiguous task, agreement rises to $\kappa = 0.87$ (raw $93.5\%$, $n=62$). The full agreement table,
every per-item disagreement, and the script that computes them are in the released artifact
(Section~\ref{sec:availability}). We designed to the Agentic Benchmark Checklist
\citep{zhu2025benchpractices}: outcome validity comes from kernel-measured properties rather than
tests or judges, and task validity from the requirement-to-check mapping plus expert review.

\section{Metrics}
Let an attempt produce properties $p$, grade $\mathrm{pass}(p) \in \{0,1\}$ (all checks within
tolerance), validity $v(p) \in \{0,1\}$ (renderable, topologically valid solid), claim
$d \in \{0,1\}$ (agent reported \textsc{done}), and confidence $c \in [0,1]$.
\begin{itemize}
\item \textbf{Success rate} $\Pr[\mathrm{pass}]$ and \textbf{success@$k$} (any of $k$ seeds
  passes), plus \textbf{pass\textasciicircum{}$k$} consistency (all of $k$ seeds pass)
  \citep{yao2024taubench}; capability and reliability are reported separately.
\item \textbf{Silent-failure rate} $= \Pr[\,v \wedge d \wedge \neg\mathrm{pass}\,]$: a valid,
  renderable solid, an explicit completion claim, and at least one failed semantic check. Loud
  failures ($\neg v$) and honest non-claims ($\neg d$) are excluded by construction, so the metric
  isolates agent-side overclaiming under a sound oracle.
\item \textbf{Completion-claim calibration}: ECE with equal-mass bins and reliability diagrams
  \citep{guo2017calibration} over $(c, \mathrm{pass})$ pairs; \textbf{AUROC} of $c$ as a
  discriminator of pass/fail (informative at low base rates
  \citep{kaddour2026overconfidence}); Brier score.
\item \textbf{Claim rate} $\Pr[d]$ and \textbf{claim precision}
  $\Pr[\mathrm{pass} \mid d]$: the architecture-neutral pair a user experiences, reported
  alongside the silent-failure rate (interference and manufacturability checks arrive with the
  assembly tier in v2 and are not reported here).
\item \textbf{Cost}: input and output tokens and tool calls per attempt.
\end{itemize}

\section{Results}
\label{sec:results}
\paragraph{Populations.} A mid-study audit (Section~\ref{sec:discussion}) found a protocol
loophole in the original tool-use loop and permissive checks in the original oracle; we hardened
both, re-executed and regraded every stored attempt, and re-ran tool-use under the fixed protocol
where budget allowed. We therefore report three populations: \textbf{primary}:
claude-haiku-4-5, claude-sonnet-4-6, deepseek-v3.2, and qwen3.7-max (four vendors), complete
13-task coverage, hardened oracle, fixed-protocol tool-use ($n{=}156$ per architecture; all
headline claims and statistics come from here); \textbf{ablation}: a tool-use arm rerun under
the legacy claim contract with identical feedback on all four models ($n{=}156$; the sonnet arm
was run through the OpenRouter gateway to the same model); and \textbf{protocol sensitivity}: the
original-protocol tool-use runs of the same four models, regraded under the hardened oracle (143
attempts; the qwen run covers 26 of its 39 cells), reported descriptively. An incomplete
exploratory run of a fifth model (gpt-5.5, an easy-tier prefix under the original protocol) is
included in the release for completeness but is not used in this paper. The per-condition
success and silent-failure rates carry Wilson 95\% intervals in the released statistics file; bracketed intervals on contrasts are task-cluster bootstraps; paired tests are exact McNemar over (task,
model, repetition) triples; a GEE logistic model with task clusters confirms each headline
contrast (with only 13 clusters its sandwich standard errors are anti-conservative, so the
cluster bootstrap is the primary interval and the GEE is corroborating). The paper tests two
pre-specified contrasts (single-shot versus each verified condition); every other comparison is
reported descriptively, without a multiplicity correction, and should be read that way. The three repetitions are resamples at provider-default decoding (no seed control);
attempts within a task are correlated, and the cluster intervals reflect that. The headline contrast holds within every repetition
taken alone: single-shot silent failures were 5, 9 and 8 of 52 per repetition against 1, 4 and 0
for enforced tool-use and 1, 2 and 1 for the checker pipeline, so no single resample drives it.

\begin{table}[t]
\centering
\small
\input{tab_main}
\caption{Primary population. \emph{Silent} = valid renderable solid $+$ explicit completion claim
$+$ $\geq$1 failed semantic check. ECE = equal-mass 5-bin expected calibration error. Counts, not just rates: silent failures are 22/156, 5/156, 4/156 by condition.}
\label{tab:main}
\end{table}

\begin{figure}[t]
\centering
\includegraphics[width=\textwidth]{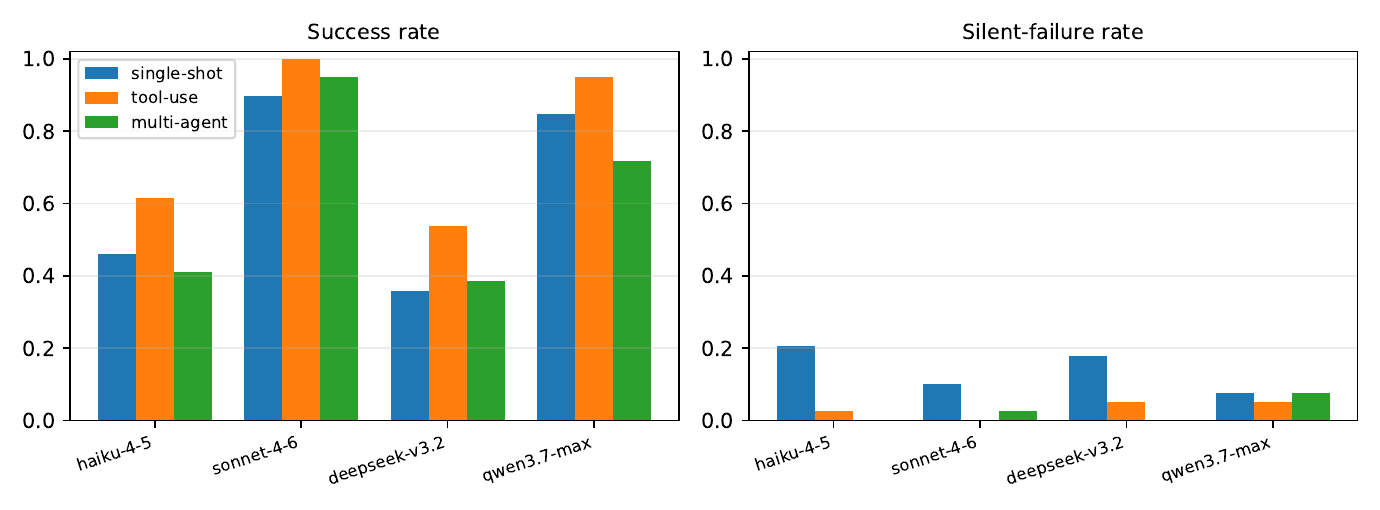}
\caption{Success and silent-failure rates by model and agent condition (primary population).}
\label{fig:main}
\end{figure}

\paragraph{F1: unverified claims are the hazard; enforced verification almost removes it.}
Table~\ref{tab:main} and Figure~\ref{fig:main} give the per-model breakdown. Single-shot agents claimed \textsc{done} on 153/156 attempts; 22 (14.1\%; task-cluster CI [5.8, 24.4], Wilson [9.5, 20.4]) were silent failures, and claim precision ($\Pr[\text{pass} \mid \text{claim}]$,
the architecture-neutral quantity a user experiences) was 64.1\%. The measure-then-claim
tool-use loop cut silent failures to 5/156 (3.2\%) against single-shot's 22/156 (McNemar over (task, model, repetition) triples, removed 18, created 1, $p{<}10^{-4}$; the margins, not the pairing, carry the meaning given uncontrolled decoding;
silent-rate difference CI [3.8, 19.2] pp), with claim precision 96.0\% and a
selective claim rate of 80.8\%. Success also rose (64.1\%$\to$77.6\%; paired McNemar fixed 25, broken 4, $p{=}1.0{\times}10^{-4}$; GEE task-clustered
$p{=}0.00014$). The single-variable ablation isolates the contract: rerunning tool-use under the legacy claim contract with \emph{identical feedback} on all four models ($n{=}156$) produced 14 silent failures against the fixed contract's 5 (Fisher $p{=}0.056$; haiku 6 vs 1, sonnet 4 vs 0, deepseek 4 vs 2, qwen 0 vs 2), claim precision of 67.6\% against 96.0\% ($p{<}10^{-9}$), and success of 64.1\% against 77.6\% ($p{=}0.013$). The silent-failure difference on its own is marginal and one model runs the other way on small counts, and the sonnet and qwen legacy arms were run three months after their fixed-protocol counterparts (the sonnet arm also through a different gateway), so those two arms carry a time and routing difference as well as the contract difference; the contract's effect on what gets claimed is not marginal. It converges with the run-level sensitivity: the original loophole runs put most tool-use silent failures at unverified claim points, and two harness variables changed between those runs, which is why the ablation cell was run at all. This sharpens, and partially retracts, the reading of our own
pilot and the optimistic framing of feedback-loop systems \citep{barkley2026cadsmith}: verification
discipline at claim time, rather than iteration, is what closes the semantic gap.

\paragraph{F2: the independent checker trades coverage for precision without improving capability.}
The planner--coder--checker pipeline also reached 95.5\% claim precision with 4/156 silent failures (McNemar vs.\ single-shot: 19 removed, 1 created, $p{<}10^{-4}$), but at a 57.1\% claim rate and with no success gain over single-shot (61.5\% vs.\ 64.1\%). Decomposing its 60 failed episodes: 53 never produced a valid solid (the coder, fed second-hand checker feedback rather than raw kernel traces, crashes more), 3 were valid-but-wrong parts correctly held back, 4 were silent failures; the checker also withheld claims on 11 episodes whose parts were actually correct. In selective-prediction terms
\citep{geifman2017selective, mozannar2020defer}, enforced tool-use and the checker pipeline sit at
different points on the same risk--coverage curve (81\% coverage at 4.0\% conditional risk vs.\ 57\% at 4.5\%); on this suite the architecturally simpler tool-use point dominates. The checker's
verdict is also a different claim-bearer than the generator's own \textsc{done}; claim precision
is the cross-architecture metric precisely because it is agnostic to who claims.

\begin{figure}[t]
\centering
\includegraphics[width=0.92\textwidth]{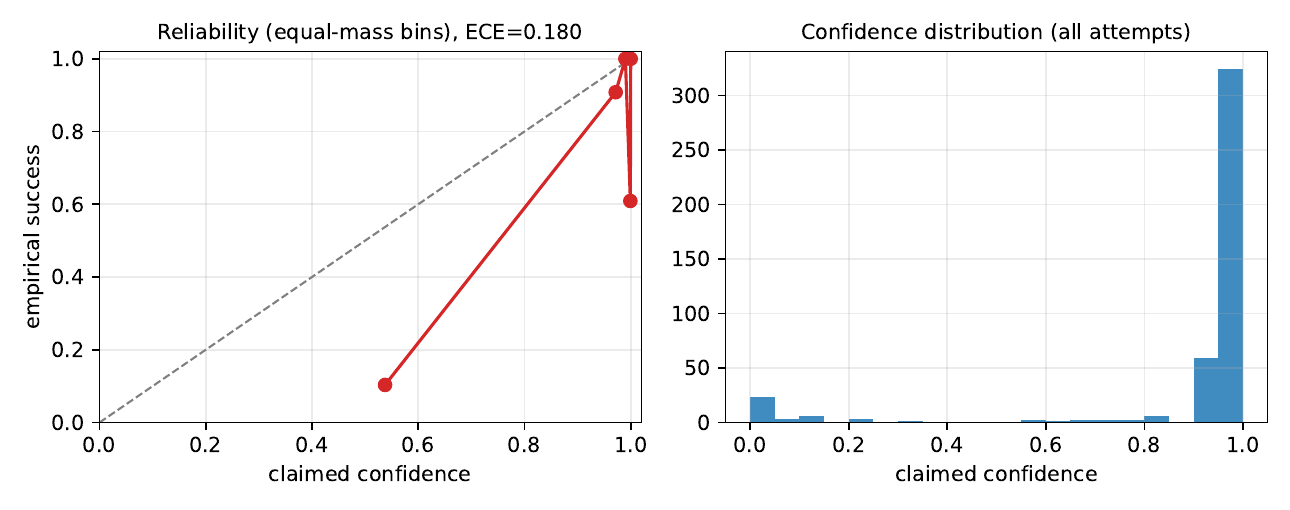}
\caption{Reliability diagram and confidence distribution, primary population, pooled over the three conditions (hence the single pooled ECE in the panel title; per-condition ECE is in F3, per-model ECE in Table~\ref{tab:main}).}
\label{fig:rel}
\end{figure}

\paragraph{F3: with verification discipline, stated confidence becomes informative.}
Single-shot confidence is saturated (claim rate 98\%, mean 0.97; Figure~\ref{fig:rel}) and poorly calibrated (ECE 0.324); its silent failures carried mean confidence 0.93 (min 0.60, pooled over all conditions). Under enforced tool-use and the checker pipeline, ECE falls to 0.091 and 0.110, and per-model AUROC (the pooling-robust quantity) rises for three of the four models (haiku 0.72$\to$0.98/0.99; sonnet 0.93$\to$1.00 under the checker on few negatives, its tool-use arm having no failures to rank; deepseek 0.65$\to$0.83/0.93). The fourth, qwen3.7-max, falls from 0.65 to chance (0.50 and 0.55): its confidence is saturated at 1.0 regardless of outcome, so verification discipline improved its claims but not its stated uncertainty, and the two are separable. The checker's confidence restricted to valid-solid episodes discriminates at AUROC 0.68 (7 negatives). We do not claim the architectures \emph{teach} calibration; the
verified conditions let the agent condition its claim on measurements, which is exactly the
point.

\paragraph{F4: context divergence tracks episode breakdown but adds nothing over episode length, and misses subtle wrongness.}
Episode-mean CDS (defined in Section~\ref{sec:method}) separates passing from failing multi-agent
episodes (0.105 vs.\ 0.518 for never-built solids; Mann--Whitney over passes versus all failures
$p{<}10^{-15}$; AUROC 0.86 as an
episode-failure predictor). That number needs its baselines: the number of coder--checker rounds
predicts the same outcome at AUROC 0.87, and the validity flag alone at 0.94, so CDS adds no
predictive value over episode length. Its interest is diagnostic rather than predictive: it names
which dimension the roles disagree about while the episode is still running. Stratifying by
validity removes even that: the seven valid-but-wrong episodes average CDS 0.107, indistinguishable
from passes (0.105). Drift between the roles' stated dimensions is an early breakdown signal, but
it does not flag the plausible-wrong parts that motivate this paper. We report this null
explicitly.

\begin{figure}[t]
\centering
\includegraphics[width=\textwidth]{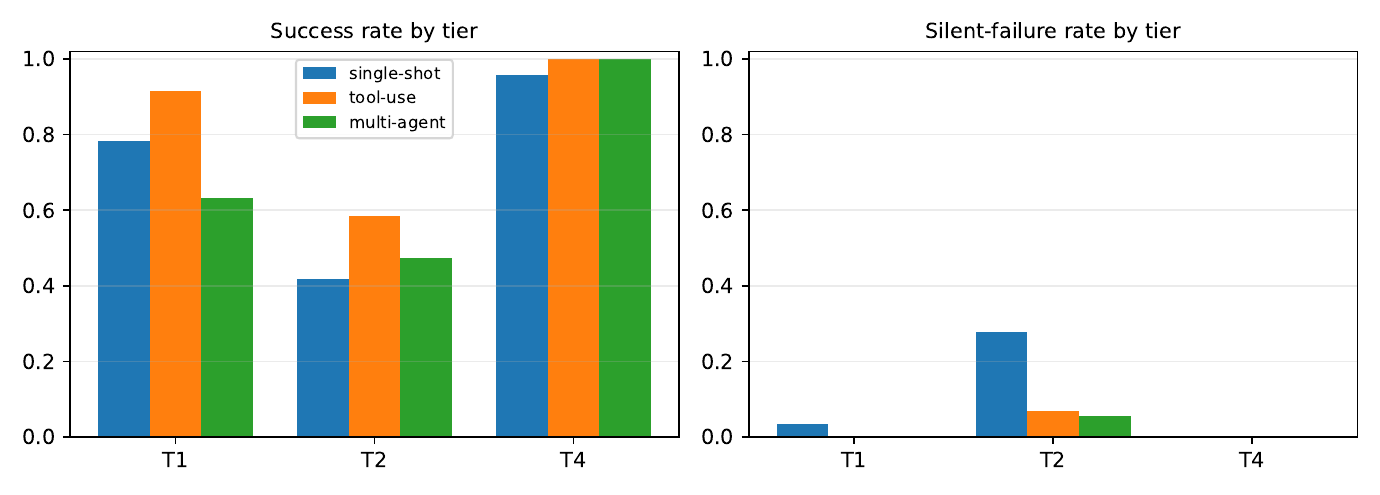}
\caption{Success and silent-failure rates by tier and condition (primary population).}
\label{fig:tier}
\end{figure}

\paragraph{F5: silent failures live in internal multi-feature geometry.}
All 31 primary silent failures are T1/T2 parts; 19 of 31 come from two tasks (the L-bracket and the counterbored plate: internal features and non-coaxial holes), and the two T4 edit tasks produced none (Figure~\ref{fig:tier}; exploratory slices, not tested).
Hazard tracks internal-feature complexity rather than headline task difficulty.
Because the L-bracket specification admits two readings of the base depth
(Section~\ref{sec:method}), we also report the primary contrast without that task: on the remaining
12 tasks ($n{=}144$ per condition) single-shot produced 16 silent failures (11.1\%), enforced
tool-use 2 (1.4\%), and the checker pipeline none, with the paired single-shot versus tool-use test
unchanged in direction (15 removed, 1 created, McNemar $p{=}0.0005$). The ambiguous task inflates
the absolute rates and narrows, rather than creates, the gap between conditions.

\paragraph{F6: the oracle audits itself, and rates are bounded on both sides.}
Prompted by an adversarial engineering review of our own suite, we hardened the oracle: axial
extents on every cylindrical check (blind holes, transposed sections), per-hole radial deviation
plus pattern-centre and angular-uniformity constraints on bolt circles, translation-invariant
mass-centre offsets (mirrored and wrong-face features), body counts, and $\pm$8\% volume bands.
Re-executing and regrading all 519 stored attempts flipped 18 verdicts pass$\to$fail and
\textbf{zero} fail$\to$pass, exposing 17 silent failures the original checks had hidden, 35\% of the 48 silent
failures in the regraded corpus. Property oracles are necessary-condition
screens: any property they do not model can only add failures, so along that axis the measured
rates are lower bounds. Grader false negatives push the other way, and the blind expert audit
(Section~\ref{sec:method}) bounds them: outside the spec-ambiguous task the expert overturned 4 of
62 verdicts, all from grader-fail to expert-pass, and 3 of those 4 had been counted as silent
failures (two single-shot counterbored plates, one tool-use stepped shaft). Reading the expert as
ground truth for those parts moves the primary rates to 20/156 (12.8\%) single-shot and 4/156
(2.6\%) tool-use, and outside that task no grader pass was overturned. On the ambiguous task itself
the expert, reading the base depth the other way, passed 9 of its 13 silent failures and failed one
part the grader passed; F5 reports the contrast without that task. The rates are therefore bounded
on both sides, and benchmark authors should expect their own graders to fail the audit they apply
to agents.

\begin{table}[t]
\centering
\small
\resizebox{\textwidth}{!}{\input{tab_tolerance}}
\caption{Tolerance-band sensitivity, primary population ($n{=}156$ per condition). Every band is
scaled by the factor in the first column; SS single-shot, TU tool-use, MA multi-agent. The
published bands reproduce the stored verdicts exactly.}
\label{tab:tolerance}
\end{table}

\paragraph{F7: the measurand is stable under tolerance-band scaling.}
Every verdict depends on the published bands, so we regraded the primary population with every
band (dimensional tolerances, cylinder radius and axial-extent tolerances, bolt-circle radial,
centre and angular tolerances, and the half-width of every range check) scaled by a common factor
(Table~\ref{tab:tolerance}). At the published bands the offline regrade reproduces all 468 stored
verdicts. Halving every band changes no verdict: every part that passes at the published bands
also passes at half of them, so the silent failures are gross errors (a missing counterbore, a
hole on the wrong pitch circle, a wall added outboard), not marginal ones. Doubling the bands
forgives two of the 22 single-shot silent failures and quadrupling forgives seven; no tool-use or
checker verdict moves at any scaling above half-width, and the single-shot minus tool-use
contrast survives every scaling from $\times$0.25 to $\times$4 (6.4 to 10.9 pp, McNemar
$p < 0.01$). Tightening to a quarter of the published bands raises silent failures in every
condition (33, 17 and 14 of 156), because parts correct to the millimetre are not correct to a
quarter-millimetre, and leaves the ordering of conditions unchanged. The conclusion does not rest
on where the bands were drawn.

\paragraph{Protocol sensitivity.}
The retained old-protocol tool-use runs serve as sensitivity data: deepseek-v3.2 under the loophole contract showed 11/39 silent failures versus 2/39 under the fixed contract, the largest per-model swing and the pattern F1 predicts for a weak self-verifier. These rows are
consistent with the primary findings but are not statistically interpreted.

\paragraph{Costs and consistency.} Mean tokens per attempt (input/output): 211/735 single-shot, 2{,}821/1{,}549 tool-use, 2{,}279/2{,}787 multi-agent: verification discipline costs roughly an
order of magnitude more input tokens than single-shot. Pooled success@3 is 81.4\% vs.\ pass\^{}3 53.8\%: best-of-$k$ reporting overstates dependable capability
\citep{yao2024taubench}.

\paragraph{Worked examples.} Three of the five fixed-protocol tool-use silent failures are
L-brackets, among them a claude-haiku-4-5 part that was valid, rendered and claimed at confidence
0.95 while failing its depth, volume and mass-centre checks. Of the two outside that task, the one
the expert audit also failed is a deepseek-v3.2 flanged bushing whose 20~mm bore stopped halfway
(15 of 30~mm): the kernel feedback reported the 15~mm bore, and the agent's final reply noted that
the bore was half its required length and still ended with \textsc{done} at confidence 0.60. In
single-shot, claude-haiku-4-5 produced a four-hole plate whose holes were 5~mm blind holes in an
8~mm plate, caught only by the hardened axial-extent check; and in the original-protocol tool-use
run, deepseek-v3.2 shipped a six-bolt flange failing bore and bolt-circle checks at
``\textsc{done}, confidence 1.0.'' Every attempt is inspectable in the released 3D viewer.

\section{Error modes}
\label{sec:errors}
Separating failure layers on the primary population: of 151 failed attempts, 95 never produced
a valid solid (execution or modeling aborts), and 56 produced valid geometry that failed semantic
checks. Among the latter only, the dominant signatures are missing or mis-sized
internal features (counterbore depth or face, bore and hole extents, step sections); a wrong
envelope dimension is also common, usually alongside a feature error or, on the L-bracket, as the
disputed base depth. We do not over-interpret multi-label counts across
layers: a crashed attempt fails every check vacuously, and earlier drafts of this paper made that
conflation; the released analysis code now reports the layers separately.

A practitioner-coded failure taxonomy with inter-rater agreement, following
\citet{cemri2025mast}, is not part of this paper; the released transcripts are the material for it
(Section~\ref{sec:discussion}, roadmap).

\section{Availability, licensing, and maintenance}
\label{sec:availability}
Everything the paper reports is derived from the released artifact at \href{\artifacturl}{\texttt{\artifacturl}}, archived
with a versioned DOI at \href{https://doi.org/\artifactdoi}{\texttt{https://doi.org/\artifactdoi}}. It contains: the 13 task specifications
(JSON) with their 122 checks and published tolerance bands; the 13 reference solutions; the grader,
the measurement extractor, the executor, and the three agent harnesses; every graded attempt as a
JSON record carrying the final code, the measured-properties dictionary, the verdict, the agent's
completion claim and stated confidence, token counts, and the per-round context-divergence scores
for multi-agent episodes. The records are the 519-attempt regraded corpus (the single-shot and
multi-agent attempts of the four primary models, their 143 original-protocol tool-use attempts,
and 64 exploratory gpt-5.5 attempts), each with its verdict under both the original and the
hardened oracle; the 156 fixed-protocol tool-use attempts and the 156 legacy-contract ablation
attempts, graded under the hardened oracle; and the 60-attempt pilot. The primary population (468)
is the 312 single-shot and multi-agent attempts of the four models plus the 156 fixed-protocol
tool-use attempts. The artifact also holds full multi-turn transcripts for the 156 fixed-protocol
tool-use attempts and the 156 ablation attempts; the expert audit behind the $\kappa$ analysis
(the sampled items with their grader verdicts, the agreement table, and every disagreement); the
analysis scripts that produce the statistics, tables, and figures in this paper (the remaining
counts in the text are direct tallies of the released records); and a static 3D viewer of every
attempt against its reference. Code is released under the MIT licence; task
specifications, reference solutions, attempt records, and transcripts under CC~BY~4.0. The
authors bear all responsibility for the released material; it contains no personal data, and the
only human judgement it records is a co-author's blind scoring of geometry. Hosting is a public
GitHub repository for the living suite and a Zenodo record for the frozen version reported here
(suite v1.1); the record cited in this paper is immutable, later suite versions are tagged
releases with their own DOIs, and corrections are logged in an errata file rather than applied
silently. The first author maintains the repository and answers issues; the release will be kept
runnable against the pinned kernel version for at least three years. A datasheet
\citep{gebru2021datasheets} and a reproducibility checklist are in the appendix.

\section{Discussion, limitations, and roadmap}
\label{sec:discussion}
\paragraph{What the results mean.} In the primary population, silent-failure rates tracked one variable: whether the agent could
say \textsc{done} without having seen measurements of its final geometry. Unverified claiming ran
at an 18.3\% conditional silent rate among valid claimed parts (22/120); the two verified
configurations sat at 4.0\% and 4.5\% (5/126 and 4/89). The original-protocol tool-use runs point
the same way, with deepseek-v3.2 the worst case (11 silent failures in 34 claims). That reframes the design question for
engineering copilots from ``which agent pattern?'' to ``where does the claim sit relative to the
measurement?'', a question of harness contract, cheap to enforce and testable, echoing the
external-feedback conclusions of the self-correction literature \citep{gou2024critic,
kamoi2024selfcorrection}. The same lesson applied to us: our first harness contained the loophole
and our first oracle the leniency that this paper warns about, and both materially moved the
numbers. We kept the audit trail in the paper because a reliability benchmark should not exempt its own
grader from the scrutiny it applies to agents.

\paragraph{Pre-specified predictions and what the data did to them.} The study plan
fixed three predictions before any agent was run, and the released plan file carries them
verbatim. The plan was not lodged with a registry, and its first commit to the repository is the
same commit as the results, so the order rests on our account. \textbf{P1}, that tool-use with kernel feedback beats single-shot on dimensional
correctness, is supported (F1), with the ablation adding the attribution to the claim contract.
\textbf{P2}, that context drift between roles raises the constraint-violation rate on multi-part
specifications (tested here on the multi-feature tier, as the suite has no assembly tier), is not
supported: divergence predicts episodes that never build a solid, and the seven
valid-but-wrong episodes are indistinguishable from passes on CDS (F4). \textbf{P3}, the plan's
headline, predicted that the silent-failure rate improves far less with model capability than
syntactic success does. The single-shot data do not support it as stated. Among valid claimed
parts, the conditional silent rate falls from 31\% (haiku, 8/26) and 37\% (deepseek, 7/19) to
10\% (sonnet, 4/39) and 8\% (qwen, 3/36) as the valid-solid rate rises from 54--67\% to
92--100\%; both improve together. What survives is narrower and, we think, the more useful
statement: the strongest single-shot model still ships one silently wrong part in ten claimed
valid parts, and the reduction to 4--5\% comes from the claim contract, not from capability (the
fixed-protocol tool-use rate is 5/126 valid claims pooled, with sonnet at 0/39 and qwen at 2/39).
Two pre-specified analysis choices also changed. The plan named a mixed-effects model with task
and seed random effects; provider-default decoding gives no seed control, so the reported models
are task-clustered GEE and cluster bootstraps, which estimate the same contrasts without a seed
term. The plan set an expert-versus-grader agreement target of $\kappa \geq 0.8$; the full
81-part sample reached 0.64, and 0.87 only after excluding the one task whose specification the
expert read differently (Section~\ref{sec:method}). We report the miss rather than the
post-hoc subset alone.

\paragraph{Would a machinist catch these anyway?} In a conventional drawing--CAM--inspection
chain, many of these specific failures would surface downstream: a bolt circle off pattern fails
first-article inspection; a blind hole is caught at assembly. Three answers. First, in
model-based-definition and direct-to-CAM workflows the model \emph{is} the spec; inspection then
verifies against the wrong geometry, and the error self-certifies. Second, even when caught, the
cost is scrap, rework, and lead time; the copilot's value proposition collapses if every output
needs full manual re-inspection, and if it does not, the engineer must judge when to rely on it,
the appropriate-reliance problem \citep{schemmer2023appropriate}. Third, the failures concentrate in
internal multi-feature geometry (F5), the class that slips visual review. We concede the converse:
many of the silent failures, most of them the disputed L-bracket base depth, also carry a wrong
envelope dimension, which downstream inspection catches reliably.

\paragraph{Tolerance policy.} The bands are \emph{model-correctness} bands, not functional fit
tolerances: a generated part 0.3\,mm off its specified bushing OD is a wrong model, not a loose
fit, and grading fit dimensions functionally (IT6--7, tens of microns) would test arithmetic
nobody asked the agent to do. Bands are published per check with the suite, and F7 shows that scaling all of them
by a factor of four in either direction does not change the paper's contrasts. A single declared
policy (ISO~2768 class $\times$ stated multiplier) would remove the residual inconsistency a
careful reader can find across tasks; it is on the roadmap, not in this release.

\paragraph{Calibration metrics under saturated confidence.} Stated confidences are concentrated at the top of the scale (single-shot mean 0.97; one model reports 1.0 regardless of outcome), so equal-mass bins collapse onto a few distinct values and the ECE figures depend on the binning; per-model AUROC is binning-free and is the quantity we lead with, with ECE reported alongside it.

\paragraph{Contamination and headroom.} The parts are canonical and abundantly represented in
training corpora; memorization should help agents be \emph{right} when they claim, so measured
silent-failure rates on these parts are, if anything, optimistic for novel geometry, the
direction that strengthens the warning but weakens absolute rates. Frontier models saturate the
suite's success axis (sonnet at 90--100\% across conditions), so the benchmark's discriminating axis is
reliability, not capability; parametric re-instantiation (schema-supported, v2) is the designed
contamination control.

\paragraph{Limitations.} The suite is 13 tasks and 122 checks. That is enough to separate
architectures with paired tests across four vendors, and not enough to rank models finely or to
estimate absolute rates that would transfer to a different task mix; silent failures concentrate
in two tasks (F5), so the mix matters. Results are a dated snapshot of pinned model versions. The primary population is four models from four vendors and includes no OpenAI or Google model (the gpt-5.5 run was incomplete). We compare generic architectures and did not run published text-to-CAD systems on the suite; the tool-use condition resembles CADSmith's kernel-measurement outer loop \citep{barkley2026cadsmith} without its documentation retrieval or VLM judge. Each condition uses one prompt, and we did not vary its wording. Three repetitions per cell at provider-default decoding bound the
variance estimates; 13 task clusters bound the cluster bootstrap. Thirteen of the 31 primary silent failures come from the one task whose specification
admits two readings; the headline contrast survives excluding it (F5), and absolute rates on that
task should be read as an upper bound on agent error. The oracle, though hardened and
audited, remains a necessary-condition screen: silent-failure rates are bounded below by
unmodelled properties and above by grader false negatives, which the expert audit puts at 3 of
the 18 silent failures outside the ambiguous task (F6). The audit has a single expert rater, who is
a co-author; a second, external rater would make it stronger. Blind
expert hand-scoring of a stratified 81-part sample bounds the residual false-positive and
orientation-driven false-negative directions and agrees with the grader at $\kappa = 0.64$
(0.87 excluding one spec-ambiguous task; Section~\ref{sec:method}). CDS measures divergence of stated
beliefs and predicts breakdown, not subtle wrongness; we report that null. The human side of
the hazard (whether engineers catch silent failures when an agent reports success) is out of
scope and planned as a follow-up study.

\paragraph{Broader impact.} The direct benefit is a cheap, adoptable harness contract:
any CAD copilot can refuse to accept a completion claim until the agent has seen kernel
measurements of its final code, and this paper measures what that buys. The benchmark and the
released attempt records also give evaluators a domain in which agent overclaiming can be
measured without a judge model. The risks run the other way. Absolute rates from 13 canonical
parts will be quoted out of context, in both directions: as evidence that agents are unsafe for
engineering work, or, because the verified conditions look clean, as evidence that a verification
loop makes them safe. Neither follows; the rates are bounded on both sides and specific to one suite, and the human side
of the hazard (whether engineers catch what the agent misses) is untested here. A part that passes
every check in this suite can still be unmanufacturable or unfit for its function; the checks are
model-correctness screens, not design validation, and nothing here should be read as certifying a
generated part.

\paragraph{Roadmap (not claimed here).} The released schema supports, and the next suite version
is planned to add: dual expert/non-expert phrasings per task (following
\citet{wang2026text2cadbench}, probing the under-specification sensitivity documented by
\citet{yuan2026procad}); parametric re-instantiation of task families as the contamination
control; a multi-part assembly tier with interference checks; ISO~2768 general-tolerance classes
per feature under one declared band policy; and the practitioner-coded failure taxonomy with
inter-rater agreement. None of these is part of this paper's evidence.

\paragraph{Conflict-of-interest disclosure.} Clive Rodrigues provides CAD-for-AI consulting to a
company in the mechanical-AI data space. This work uses only open kernels and public data; no
proprietary data, tooling, or funding from that engagement was used.

\paragraph{Author contributions (CRediT).} Carson Rodrigues: conceptualization, software (harness,
auto-grader, eval infrastructure), formal analysis, visualization, writing. Clive Rodrigues:
conceptualization (engineering requirements), resources (task suite and ground-truth specifications), and investigation (failure-mode taxonomy). Aravind Reddy G: validation (blind expert pass/fail scoring and expert-versus-grader
$\kappa$), and resources (Siemens NX STEP reproduction of the reference parts).

\bibliographystyle{unsrtnat}
\bibliography{verified}

\appendix
\section{Datasheet}
\label{app:datasheet}
Following \citet{gebru2021datasheets}; the released repository carries the full version.

\paragraph{Motivation.} Created to measure, under an exact geometric oracle, how often LLM agents
that write parametric CAD ship a valid, plausible, wrong part while claiming completion, and how
well their stated confidence tracks correctness. Built by the authors with no external funding.

\paragraph{Composition.} \sloppy 13 task specifications (5 single-feature, 6 multi-feature, 2
edit-under-constraint) with 122 machine-checkable requirements (58 dimensional \texttt{approx},
13 \texttt{range}, 13 \texttt{equal}, 13 validity, 22 \texttt{cylinder\_count} with axial
extents, 3 \texttt{bolt\_circle}); 13 reference solutions; the attempt corpus, each attempt a
JSON record with task, model, condition, repetition, final code, measured properties, verdict,
failed checks, completion claim, stated confidence, steps, token counts, and per-round
context-divergence scores where applicable: 519 regraded attempts with verdicts under both oracle
versions (the four primary models, plus an incomplete exploratory fifth model that the paper does
not use), 156 fixed-protocol tool-use attempts, 156 ablation attempts, and 60 pilot attempts;
multi-turn transcripts for the 312 fixed-protocol tool-use and ablation attempts; the 81-item
expert audit (sampled items with grader verdicts, the agreement table, and every disagreement).
No personal data; no data about people.

\paragraph{Collection.} Specifications were written by the authors, one a practising mechanical
engineer, and every requirement was mapped to a check before any agent run. Attempts were
generated in June 2026 (the sonnet and qwen ablation arms in September 2026) through the vendors'
public APIs at provider-default decoding with three repetitions per (task, model, condition) cell;
tool-use episodes were capped at four model turns plus one confirmation turn, multi-agent episodes
at a plan and three coder--checker rounds, and the kernel executor at 60~s per execution. Every
attempt was graded, or re-executed and regraded, under the hardened oracle.

\paragraph{Uses.} Evaluating CAD-writing agents for silent failure and claim calibration;
testing harness contracts; studying grader leniency. Not for training text-to-CAD models on the
reference solutions if the suite is to remain an evaluation.

\paragraph{Distribution and maintenance.} Section~\ref{sec:availability}.

\input{appendix_prompts}

\section{Reproducibility checklist}
\label{app:checklist}
\begin{itemize}
\item \textbf{Code and data:} all released (Section~\ref{sec:availability}); one script regrades
  every stored attempt offline, and the released analysis scripts produce the statistics, tables,
  and figures (the remaining counts in the text are direct tallies of the released records).
\item \textbf{Models:} API identifiers exactly as called: \texttt{claude-haiku-4-5} and
  \texttt{claude-sonnet-4-6} (Anthropic API; the sonnet ablation arm as
  \texttt{anthropic/claude-sonnet-4-6} via OpenRouter), \texttt{deepseek/deepseek-v3.2} and
  \texttt{qwen/qwen3.7-max} (OpenRouter), all called in June 2026 except the sonnet and qwen
  ablation arms (September 2026) (an incomplete gpt-5.5 run ships in the release,
  unused here);
  provider-default decoding, 4,096 output tokens, three repetitions per cell, no seed control.
\item \textbf{Harness:} prompts and parsing rules verbatim in Appendix~\ref{app:prompts}; tool-use capped at four model turns plus one confirmation turn, multi-agent at a plan and three coder--checker rounds; executor timeout 60~s;
  measure-then-claim contract as defined in Section~\ref{sec:method}; the legacy contract is
  retained as the ablation arm.
\item \textbf{Statistics:} Wilson intervals on the per-condition success and silent-failure rates; task-cluster bootstrap (4,000 resamples) on
  contrasts; exact McNemar over (task, model, repetition) triples; task-clustered GEE; Fisher
  exact for the 156-vs-156 ablation; all computed by the released script and reproduced from the
  raw records before submission.
\item \textbf{Grader validation:} calibration gate on reference solutions; blind expert scoring
  of 81 parts with the answer key, every disagreement, and the $\kappa$ computation released; oracle-hardening audit with
  verdict flips reported in both directions.
\item \textbf{Pre-specified predictions:} the study plan with predictions P1--P3 is in the
  repository (not lodged with a registry); Section~\ref{sec:discussion} reports which held.
\item \textbf{Compute and cost:} the whole corpus is API calls at a few thousand tokens per
  attempt; no GPU. Regrading and the tolerance sweep run on a laptop in under an hour.
\end{itemize}

\end{document}

%% file: tab_main.tex
\begin{tabular}{llrrrrrr}
\toprule
Model & Condition & $n$ & Success & Silent & Conf. & ECE & Steps \\
\midrule
claude-haiku-4-5 & single-shot & 39 & 46\% & 21\% & 0.93 & 0.498 & 1.0 \\
claude-haiku-4-5 & tool-use & 39 & 62\% & 3\% & 0.93 & 0.102 & 3.0 \\
claude-haiku-4-5 & multi-agent & 39 & 41\% & 0\% & 0.48 & 0.061 & 2.6 \\
\addlinespace
claude-sonnet-4-6 & single-shot & 39 & 90\% & 10\% & 0.95 & 0.084 & 1.0 \\
claude-sonnet-4-6 & tool-use & 39 & 100\% & 0\% & 0.98 & 0.016 & 2.2 \\
claude-sonnet-4-6 & multi-agent & 39 & 95\% & 3\% & 0.96 & 0.030 & 1.1 \\
\addlinespace
deepseek-v3.2 & single-shot & 39 & 36\% & 18\% & 0.99 & 0.671 & 1.0 \\
deepseek-v3.2 & tool-use & 39 & 54\% & 5\% & 0.98 & 0.279 & 3.3 \\
deepseek-v3.2 & multi-agent & 39 & 38\% & 0\% & 0.56 & 0.195 & 2.8 \\
\addlinespace
qwen3.7-max & single-shot & 39 & 85\% & 8\% & 1.00 & 0.149 & 1.0 \\
qwen3.7-max & tool-use & 39 & 95\% & 5\% & 1.00 & 0.051 & 2.3 \\
qwen3.7-max & multi-agent & 39 & 72\% & 8\% & 0.97 & 0.252 & 1.7 \\
\bottomrule
\end{tabular}

%% file: tab_tolerance.tex
\begin{tabular}{lrrrrrr}
\toprule
Band scale & SS silent & TU silent & MA silent & SS success & TU success & SS$-$TU (pp), McNemar $p$ \\
\midrule
$\times$0.25 (tight) & 33/156 (21.1\%) & 17/156 (10.9\%) & 14/156 (9.0\%) & 57.0\% & 69.9\% & 10.3, $<$0.001 \\
$\times$0.5 & 22/156 (14.1\%) & 5/156 (3.2\%) & 4/156 (2.6\%) & 64.1\% & 77.6\% & 10.9, $<$0.001 \\
$\times$1 (published) & 22/156 (14.1\%) & 5/156 (3.2\%) & 4/156 (2.6\%) & 64.1\% & 77.6\% & 10.9, $<$0.001 \\
$\times$2 & 20/156 (12.8\%) & 5/156 (3.2\%) & 4/156 (2.6\%) & 65.4\% & 77.6\% & 9.6, $<$0.001 \\
$\times$4 (loose) & 15/156 (9.6\%) & 5/156 (3.2\%) & 4/156 (2.6\%) & 68.6\% & 77.6\% & 6.4, 0.006 \\
\bottomrule
\end{tabular}

%% file: appendix_prompts.tex
\section{Prompts and claim-parsing rules}
\label{app:prompts}
Verbatim from the released harness (\texttt{benchmark/agents.py}). Every first user message is the task specification, as \texttt{Specification (task <id>): <text> Units: <units>.}; the tool-use condition appends the instruction below to it, and every multi-agent role receives the specification at the start of its message.

\paragraph{System prompt, single-shot and tool-use conditions.}
\begin{quote}\small\ttfamily\raggedright
You are a senior mechanical CAD engineer. You build parts as parametric code with the \textasciigrave{}build123d\textasciigrave{} Python library. Always assign the final solid to a top-level variable named \textasciigrave{}result\textasciigrave{}. Import everything you use. Work in millimetres, and model the part in the conventional orientation: primary axis / thickness along +Z. Reply with exactly one \textasciigrave{}\textasciigrave{}\textasciigrave{}python code block. When you are confident the part satisfies the specification, also write \textemdash{} after and outside the code block, on their own lines \textemdash{} \textasciigrave{}DONE\textasciigrave{} and \textasciigrave{}CONFIDENCE: \textless{}0..1\textgreater{}\textasciigrave{}, your calibrated probability that every requirement is met.
\end{quote}
\paragraph{Tool-use instruction (appended to the specification).}
\begin{quote}\small\ttfamily\raggedright
You may iterate: I will run your code and report measured geometry. Declare DONE only after reviewing the measurements of your latest code.
\end{quote}
\paragraph{Tool-use feedback template (one message per executed revision).}
\begin{quote}\small\ttfamily\raggedright
Kernel-measured properties of your solid:\\
\textless{}JSON with valid\_solid, bbox, volume, com, solid\_count, cylinders\textgreater{}\\
If this meets the spec, reply DONE + CONFIDENCE. Otherwise send a corrected \textasciigrave{}\textasciigrave{}\textasciigrave{}python block.
\end{quote}
\paragraph{Message when DONE accompanies unmeasured code (measure-then-claim contract).}
\begin{quote}\small\ttfamily\raggedright
Your DONE was not accepted yet: review the measurements above, then reply DONE + CONFIDENCE (no code) if the part meets the spec, or send a corrected \textasciigrave{}\textasciigrave{}\textasciigrave{}python block.
\end{quote}
\paragraph{Planner system prompt.}
\begin{quote}\small\ttfamily\raggedright
You are the PLANNER in a mechanical CAD team. From the specification, produce a short numbered build plan (features, order of operations, datums) for a \textasciigrave{}build123d\textasciigrave{} Python implementation. Work in millimetres. Do NOT write code. End your reply with one line \textasciigrave{}DIMS: \{"name": value, ...\}\textasciigrave{} \textemdash{} a flat JSON object naming every governing dimension in mm (numeric values only).
\end{quote}
\paragraph{Coder system prompt.}
\begin{quote}\small\ttfamily\raggedright
You are the CODER in a mechanical CAD team. Implement the plan with the \textasciigrave{}build123d\textasciigrave{} Python library. Always assign the final solid to a top-level variable named \textasciigrave{}result\textasciigrave{}. Import everything you use. Work in millimetres. Reply with exactly one \textasciigrave{}\textasciigrave{}\textasciigrave{}python code block, then one line \textasciigrave{}DIMS: \{"name": value, ...\}\textasciigrave{} \textemdash{} a flat JSON object of the governing dimensions (mm) your code implements.
\end{quote}
\paragraph{Checker system prompt.}
\begin{quote}\small\ttfamily\raggedright
You are the CHECKER in a mechanical CAD team. You receive the original specification and kernel-measured properties of the part the coder built. Judge whether the part satisfies every requirement. Reply with: \textasciigrave{}VERDICT: PASS\textasciigrave{} or \textasciigrave{}VERDICT: REVISE\textasciigrave{}, a one-line reason (if REVISE, say exactly what to change), one line \textasciigrave{}DIMS: \{"name": value, ...\}\textasciigrave{} \textemdash{} the governing dimensions (mm) you believe the measured part actually has \textemdash{} and one line \textasciigrave{}CONFIDENCE: \textless{}0..1\textgreater{}\textasciigrave{}, your calibrated probability that every requirement is met.
\end{quote}
\paragraph{Parsing rules.} A completion claim counts only when the token \texttt{DONE} starts its own line outside any code block (so ``this is NOT DONE'' never counts); under the measure-then-claim contract a \texttt{DONE} that accompanies fresh, unmeasured code is rejected with the message above and the episode continues. Confidence is read from a \texttt{CONFIDENCE: <0..1>} line; a missing or unparsable value is recorded as no stated confidence and excluded from calibration statistics. The checker's verdict is read from \texttt{VERDICT: PASS|REVISE}, and each role's stated dimensions from a \texttt{DIMS: \{...\}} line of flat JSON in millimetres, which is what the Context Divergence Score compares. Tool-use episodes are capped at four model turns; an episode that reaches the cap gets one confirmation turn with the measurements of its final code, in which a \texttt{DONE} counts and new code ends the episode unclaimed. Multi-agent episodes are capped at one plan and three coder--checker rounds. The kernel executor is capped at 60~s per execution.